\pdfoutput=1

\documentclass[prl,letterpaper,twocolumn,10pt,aps,notitlepage,nofootinbib,nobibnotes]{revtex4-1}

\usepackage{amsfonts}
\usepackage{amsmath}
\usepackage{bm,graphicx}
\usepackage{hyperref}
\usepackage{epsfig}
\usepackage{verbatim}   
\usepackage[subfigure]{graphfig}
\usepackage{epstopdf}
\usepackage{dcolumn}
\usepackage{color}
\usepackage{tabularx}
\usepackage{braket}
\usepackage{float}
\usepackage{tikz}
\usepackage[percent]{overpic}
\graphicspath{{./figures/}}

\def\emb{\color{blue}}
\def\emr{\color{red}}

\hypersetup{colorlinks, linkcolor = [rgb]{0,0.0,0.75}, citecolor = [rgb]{0,0.0,0.75}, urlcolor = [rgb]{0,0.0,0.75}}

\usepackage[utf8]{inputenc}
\def\be{\begin{equation}}
\def\ee{\end{equation}}
\def\bea{\begin{eqnarray}}
\def\eea{\end{eqnarray}}
\newcommand{\Tr}{\mathrm {Tr}}

\newcommand{\nn}{\nonumber}
\newcommand{\la}{\label}

\definecolor{green}{rgb}{0,.5,0}

\begin{document}

\title{Strange Quark Magnetic Moment of the Nucleon at the Physical Point }

\author{{Raza Sabbir Sufian$^{1}$, Yi-Bo Yang$^{1}$, Andrei Alexandru$^{2}$,  Terrence Draper$^{1}$,  Jian Liang$^{1}$, and Keh-Fei Liu$^{1}$ }
\vspace*{-0.5cm}
\begin{center}
\large{
\vspace*{0.4cm}
\includegraphics[scale=0.15]{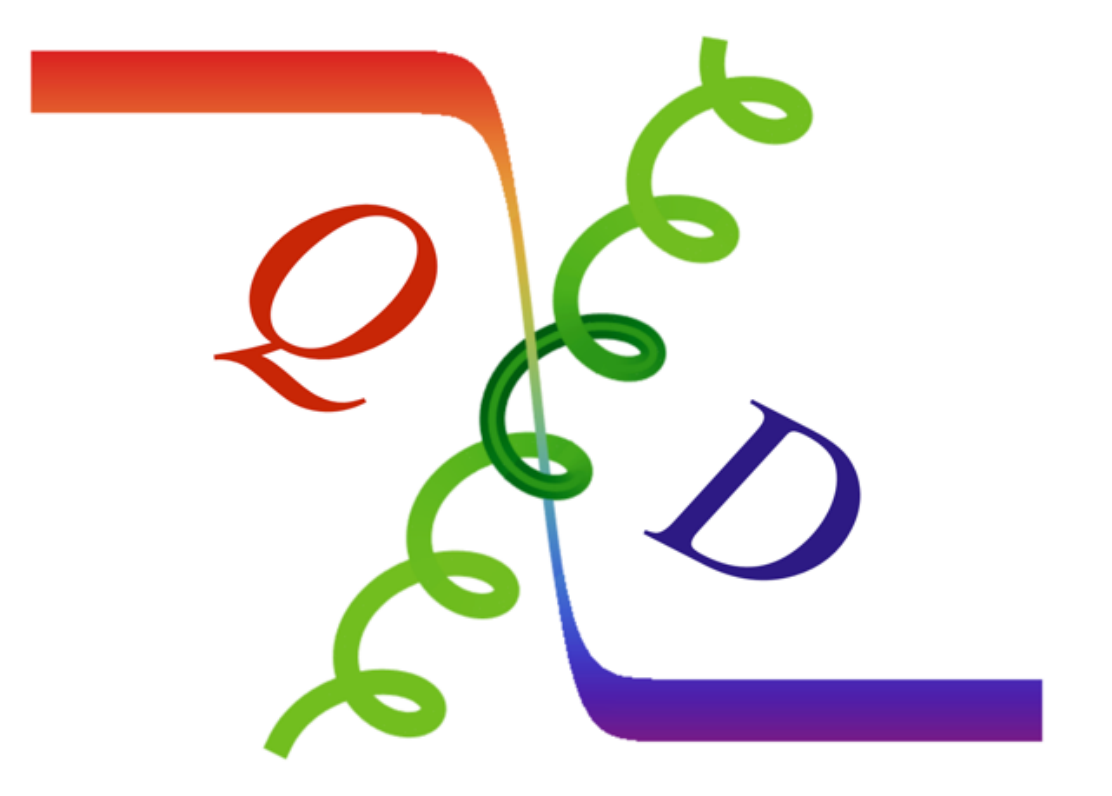}\\
\vspace*{0.4cm}
($\chi$QCD Collaboration)
}
\end{center}
}

\affiliation{
$^{1}$\mbox{Department of Physics and Astronomy, University of Kentucky, Lexington, Kentucky 40506, USA}\\
$^{2}$\mbox{Department of Physics, The George Washington University, Washington, D.C. 20052, USA}
}
%\date{\today}

\begin{abstract}
We report a lattice QCD calculation of the strange quark contribution to the nucleon's magnetic moment and charge radius. This analysis presents the first direct determination of strange electromagnetic form factors including at the physical pion mass. We perform a model-independent extraction of the strange magnetic moment and the strange charge radius from the electromagnetic form factors in the momentum transfer range of $0.051 \,\text{GeV}^2 \lesssim Q^2 \lesssim 1.31 \,\text{GeV}^2 $. The finite lattice spacing and finite volume corrections are included in a global fit with $24$ valence quark masses on four lattices with different lattice spacings, different volumes, and four sea quark masses including one at the physical pion mass. We obtain the strange magnetic moment $G^s_M(0) = - 0.064(14)(09)\, \mu_N$. The four-sigma precision in statistics is achieved partly due to low-mode averaging of the quark loop and low-mode substitution to improve the statistics of the nucleon propagator. We also obtain the strange charge radius \mbox{$\langle r_s^2\rangle_E = -0.0043 (16)(14)\,$ $\text{fm}^2$}. 
\end{abstract}

\maketitle
The determination of the strange ($s$) quark contribution to nucleon electromagnetic (EM) form factors is of immense importance since this is a pure sea quark effect. A nonzero value of the strange Sachs electric form factor (FF) $G^s_E$ at any $Q^2\neq0$ would mean that the spatial distributions of $s$ and $\bar{s}$ quarks are not the same in the nucleon. Since the extraction of the vector strange matrix elements $\langle N \vert \bar{s}\gamma_\mu s\vert N\rangle$ was proposed in Refs.~\cite{Kaplan, Mckeown, Beck} via parity-violating $e-N$ scattering for which the dominant contribution arises from interference between photon ($\gamma$) and weak boson ($Z$) exchanges by the following relation assuming isospin symmetry:
\bea \label{1}
G^{Z,p}_{E,M} (Q^2) &=& (\frac{1}{4}-\sin^2 \theta_W) G^{\gamma , p}_{E,M}(Q^2) -\frac{1}{4} G^{\gamma , n}_{E,M}(Q^2)\nn \\
 &&-\frac{1}{4} G^{s, p}_{E,M}(Q^2)\, , 
\eea
a considerable number of experimental efforts by the SAMPLE, HAPPEX, G0, and A4~\cite{Spayde:2003nr, Beise:2004py,
Aniol:2004hp, Maas:2004ta, Maas:2004dh, Aniol:2005zf, Aniol:2005zg,
Armstrong:2005hs, Acha:2006my, Androic:2009aa, Baunack:2009gy, Ahmed:2011vp} Collaborations have been going on for the past two decades. The world data constrains that $G^s_M(0)$ contributes less than $6\%$ and $\langle r_s^2\rangle_E$ contributes less than $5\%$ to the magnetic moment and the mean-square charge radius of the proton respectively~\cite{Armstrong2}. However, all these experimental results are limited by rather sizable error bars. Three different global analyses give $G^s_{M}(Q^2= 0.1\, \rm{(GeV/c)}^2)$ consistent with zero within uncertainties and differ in sign in their central values~\cite{JLiu,Gonzalez,Young}.

Despite tremendous theoretical efforts, e.g.~\cite{Jaffe, Musolf, Koepf2, Park}, a detailed convincing understanding about the sign and magnitude of strange EM FFs is still lacking. A detailed review of these theoretical efforts can be found in~\cite{Beck2}. 

\begin{figure}
\includegraphics[height=3.0cm,width=3.5cm]{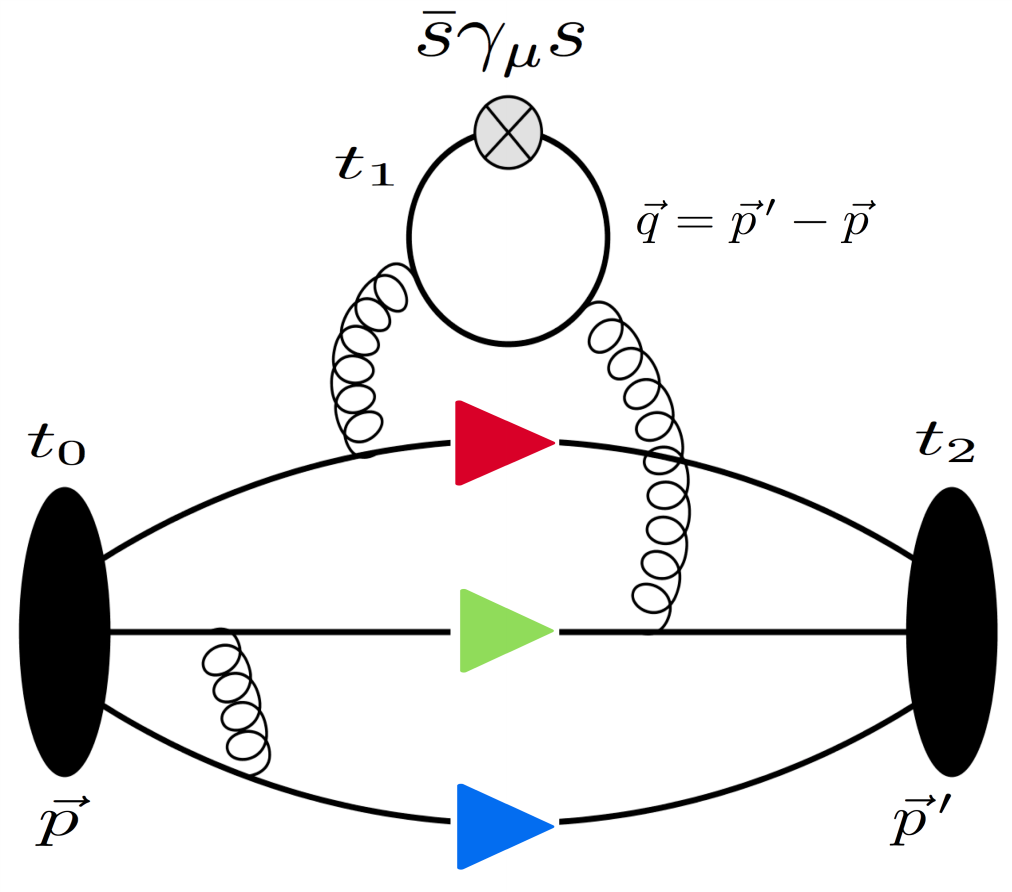}
\setlength\abovecaptionskip{-3pt}
\setlength\belowcaptionskip{-5pt}
\caption{\label{fig:DI}Disconnected three-point insertion (DI) to calculate the $\bar{s}\gamma_\mu s $ matrix element in the nucleon state }
\end{figure}

Since the direct calculation of the $s$-quark loop in the disconnected insertion (DI) is difficult and noisy in lattice QCD, there have been numerous indirect calculations to predict the strange vector FFs. Most of the calculations rely on different models (such as the heavy baryon chiral perturbation theory) or a combination of experimental and lattice QCD data of connected $u$- and $d$-quark contributions~\cite{Leinweber2, Leinweber4, Kubis}, etc.. The most recent result of such calculations has found $G^s_M(0)=-0.07(3)\mu_N$ and $G_E^s(0)$ consistent with zero~\cite {Shanahan:2014tja}. While the authors performed a linear extrapolation of $G^s_M(Q^2)$ to obtain $G^s_M(0)$, this linear behavior is different from what we observe in this work and the most recent lattice QCD analysis in Ref.~\cite{Green}.\\
\indent
 The first lattice QCD calculation was performed in the quenched approximation~\cite{Liu} and a $2+1$ flavor dynamical fermion calculation~\cite{Doi} with relatively heavy pion masses followed from the same group who obtained $G^s_M(0)=-0.017(25)(07)\mu_N$ and $G_E^s(0)$ consistent with zero. A recent lattice QCD calculation~\cite{Green} has been done with quark masses corresponding to $m_\pi = 317$ MeV and the authors obtained $G^s_M(0)= -0.022(8) \:\mu_N$ and, for the first time, a nonzero signal for $G^s_E(Q^2)$ which gave $\langle r_s^2\rangle_E= -0.0067(25)\text{ fm}^2$. However, one still has to perform the calculation at the physical pion mass and on several lattices to consider volume and finite cutoff corrections and over all beat down the noise to obtain a convincing result which will substantially sharpen our picture of strange quark contributions to the nucleon's EM structure.\\
 \indent
 Conventionally, we omit the unit nucleon magneton $\mu_N$ for $G^s_M$ in the rest of the letter. To calculate $\langle N \vert \bar{s}\gamma_\mu s\vert N\rangle$, we compute the DI on the lattice where quark loops in the nucleon sea are connected to the valence quarks through the fluctuating gauge background as shown in Fig.~\ref{fig:DI}.
We present lattice calculations of the strange EM FFs using the overlap fermion on the $(2+1)$ flavor RBC/UKQCD domain wall fermion (DWF) gauge configurations. Details of these ensembles are listed in Table~\ref{table:r0}. We use 24 valence quark masses in total for the 24I, 32I, 48I, and 32ID ensembles representing pion masses in the range $m_{\pi}\in$(135, 400) MeV to explore the quark-mass dependence of the $s$-quark FFs. 
\begin{table}[htbp]
\begin{center}
\setlength\abovecaptionskip{-3pt}
\setlength\belowcaptionskip{-5pt}
\caption{\label{table:r0} The parameters for the DWF configurations: spatial or temporal size, lattice spacing \cite{Aoki,Blum}, the sea strange quark mass under the $\overline{\text{MS}}$ scheme at {2 GeV}, the pion mass corresponding to the degenerate light sea quark mass and the numbers of configurations used in this work.}
\begin{tabular}{cccccc}
\hline
Ensemble & $L^3\times T$  &$\mathnormal{a}$ (fm) & $m_s^{(s)}$(MeV) &  {$m_{\pi}$} (MeV)  & $N_\text{config} $ \\
\hline
24I~\cite{Aoki} & $24^3\times 64$& 0.1105(3) &120   &330  & 203    \\
32I~\cite{Aoki} &$32^3\times 64$& 0.0828(3) & 110   &300 & 309 \\
48I~\cite{Blum} &$48^3\times 96$& 0.1141(2) & 94.9   &139 & 81 \\
32ID~\cite{Blum} &$32^3\times 64$& 0.1431(7) & 89.4& 171 & 200\\
\hline
\end{tabular}
\end{center}
\end{table}
We employ eigenmode deflation in the inversion of the quark matrix and use the smeared-grid $Z(3)$-noise source with low-mode substitution (LMS) to improve statistics by a very significant amount, the details of which can be found in Ref.~\cite{Ming, Anyi,Yibo}. Nucleon two-point (2pt) and three-point (3pt) correlation functions are defined as
 \bea
  && \Pi^{2pt}(\vec{p}\,',\!t_2;\!t_0)\!=\!\sum_{\vec{x}}\!e^{-i\vec{p}\,'\cdot\vec{x}} \!\Bra{0}\!T[\chi(\vec{x},\!t_2) \!\sum_{x_i\!\in G}\!\bar{\chi}_{S}(x_i,\!t_0)]\! \Ket{0}, \nn \\
&& \Pi^{3pt}_{V_\mu} (\vec{p}\,'\!, t_2;\! \vec{q},\! t_1;\!t_0) \!=\! \sum_{\vec{x}_2, \vec{x}_1}\! e^{-i\vec{p}\,'\cdot\vec{x}_2+i\vec{q}\cdot\vec{x}_1}\! \Bra{0}T[\chi(\vec{x}_2,\!t_2)\nn \\
&&\qquad \qquad \qquad \qquad \, V_\mu(\vec{x}_1,\!t_1)\!\sum_{x_i\in G}\!\bar{\chi}_{S}(x_i,\!t_0)] \!\Ket{0} ,
 \eea
where $t_0$ and $t_2$ are the source and sink temporal positions, respectively, $\vec{p},\, \vec{p}\,'$ are the source and sink momenta, respectively, $t_1$ is the time at which the bilinear operator $V_\mu(x)=\bar{s}(x)\gamma_\mu s(x)$ is inserted, $x_i$ are points on the spatial grid $G$, $\chi$ is the usual nucleon point interpolation field and $\bar{\chi}_{S}$ is the nucleon interpolation field with grid-smeared $Z_3$-noise source, and the three-momentum transfer is $\vec{q}=\vec{p}\,'-\vec{p}$ as shown in Fig.~\ref{fig:DI}. For the point sink and smeared source with $t_0=0$ and $\vec{p}=\vec{0}$ and $\vec{q}=\vec{p}\,'$ the Sachs FFs can be obtained by the ratio of a combination of 3pt and 2pt correlations with appropriate kinematic factors,
\bea \la{RatioEq}
R_\mu(\vec{q},t_2,\! t_1\!) \!=\! \frac{\Tr [\Gamma_m \Pi^{3pt}_{V_\mu} (\vec{q},t_2, \!t_1\!  ) ]} {\Tr [\Gamma_e \Pi^{2pt} (\vec{0},t_2)]} e^{(E_q-m)\cdot(t_2\!-\!t_1)}\! \frac{2E_q}{E_q+m}. \nn \\
\eea
Here, $E_q=\sqrt{m_N^2+\vec{q}\,^2}$ and $m_N$ is the nucleon mass. The choice of the projection operator for the magnetic form factor is $\Gamma_m \!=\!\Gamma_k\! =\! -i(1+\gamma_4)\gamma_k\gamma_5/2$ with $k\!=\!1,2,3$ and that for the electric form factor is $\Gamma_e\!=\!(1+\gamma_4)/2$. Then in the limit $(t_2 - t_1) \gg 1/\Delta m$ and $t_1 \gg 1/\Delta m$, we can obtain two Sachs FFs by an appropriate choice of projection operators and current directions $\mu$:
\bea
R_{\mu = i} (\Gamma_k) \xrightarrow{(t_2 - t_1) \gg 1/\Delta m, t_1 \gg 1/\Delta m} &&  \frac{\epsilon_{ijk}q_j}{E_q+m_N} G^s_M(Q^2),\nn \\
R_{\mu = 4} (\Gamma_e) \xrightarrow{(t_2 - t_1) \gg 1/\Delta m, t_1 \gg 1/\Delta m}&& G^s_E(Q^2),
\eea
with ${i,j,k}\neq4$ and $\Delta m$ the mass gap between the ground state and the first excited state. We note that $R_{\mu}$ contains a ratio $Z_P(q)/Z_P(0)$, where $Z_P(q)$ is the wave function overlap for the point sink with momentum $\vert\vec{q}\vert$. It is unity in the continuum limit, but has a small $q^2 a^2$ error at finite lattice spacing. We checked this ratio for the 32I (smallest $a$) and the 32ID (largest $a$) lattices, found its effect on the extrapolated magnetic moment and charge radius is only about $1\%-2\%$ and thus ignored it.

We incorporate a global-fit technique described in Ref.~\cite{Yibo2} to determine the $s$-quark mass by matching to the renormalized $s$-quark mass at the 2\,GeV scale in the $\overline{\text{MS}}$ scheme and use normalized vector currents~\cite{Zhaofeng}. 
To control the excited-state contamination and obtain better signal-to-noise ratios we perform a joint two-state correlated fit by simultaneously fitting the standard 3pt/2pt ratio $R(t_2,t_1)$ and the widely used summed ratio $SR(t_2)$~\cite{Maiani} to calculate DI matrix elements. We call this hybrid method the combined fit (CF) throughout the rest of this work. For more details, see Ref.~\cite{Yibo}. The $R(t_2, t_1)$ and $SR(t_2)$ fitting formulas for a given direction of current and momentum transfer can be written, respectively, as
\bea \la{CF-method}
&&R(t_2,t_1) = C_0 + C_1 e^{-\Delta m(t_2-t_1)} +C_2 e^{-\Delta m t_1} +C_3 e^{-\Delta m t_2}, \nn
\eea
\bea
&&SR(t_2) =  \sum_{t_1 \geq t'}^{t_1\leq (t_2-t'')} R(t_2,t_1)\nn \\
&&=(t_2-t' - t^{''} +1)C_0 + C_1 \frac{e^{-\Delta m t''}-e^{-\Delta m (t_2-t' + 1)}}{1-e^{-\Delta m }} \nn \\
&&\!+\! C_2 \frac{e^{-\Delta m t'}\!-\!e^{-\Delta m (t_2-\!t'' \!+\! 1)}}{1-e^{-\Delta m }}\!+\! C_3(t_2-\!t' -\!t''+\!1)e^{-\Delta m t_2}. \nn
\eea
Here, $t'$ and $t^{''}$ are the number of time slices we drop at the source and sink sides, respectively, and we choose $t'=t''=1$. $C_i$ and $\Delta m$ are fit parameters. The present scheme with the CF technique allows us to obtain a stable fit and control the excited-state contamination. We find, for the lighter quark masses on the 24I and 32I ensembles, the enhancement in the signal-to-noise ratio is approximately $5\%-10\%$ and near $m_\pi=140$ MeV for the 48I and 32ID ensembles the CF fit is more stable compared to the $SR$ and $R$ methods separately. 

In Fig.~\ref{fig:Ratio}, we present the result of CF for a particular case, the 48I ensemble with quark masses for the nucleon corresponding to $m_\pi = 207\, \text{MeV}$, $Q^2=0.0515\, \text{GeV}^2$, and several source to sink separations $t_2\in[5-9]$. We show the $SR(t_2)$ plot with an inset in the $R(t_2, t_1)$ plot. One can clearly see from the $SR$ plot that the slope is negative and from the $R$ plot that the 3pt/2pt ratio saturates near $t_2 =9$. The orange and cyan bands in the $R$- and $SR$-plots show the error bound obtained from the CF, which is $G^s_M(Q^2=0.0515\, \text{GeV}^2)=-0.029(9)$. We present this plot, in particular, to show how one can obtain a reliable and stable fit near the physical $m_\pi$. 
\begin{figure} \centering
\begin{tikzpicture}[      
        every node/.style={anchor=south west,inner sep=0pt},
        x=1mm, y=1mm,
      ]   
     \node (fig1) at (0,0)
       {\includegraphics[height=4.15cm,width=7.0cm]{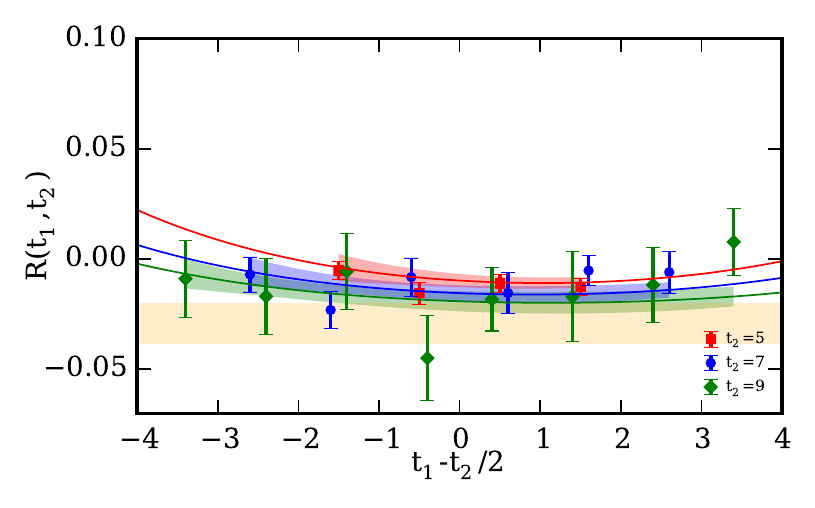}};
     \node (fig2) at (17,19)
       {\includegraphics[height=2.0cm, width=4.0cm]{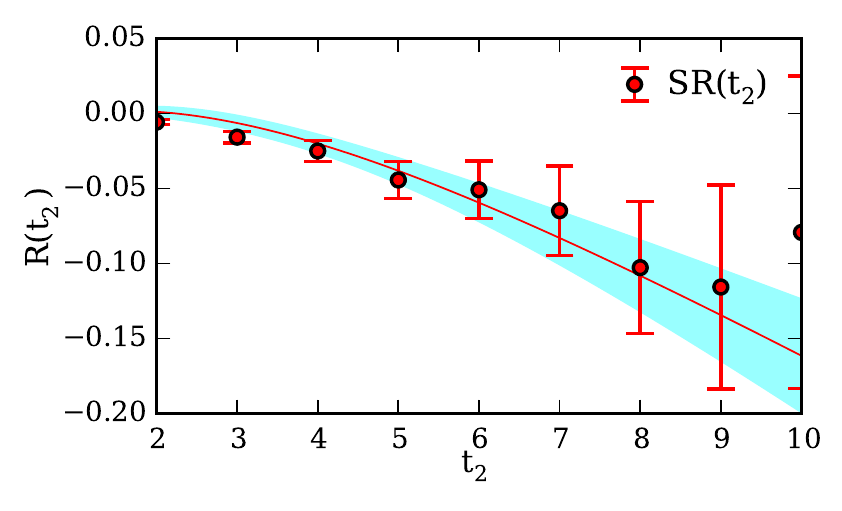}};
\end{tikzpicture}
\setlength\abovecaptionskip{-5pt}
\setlength\belowcaptionskip{-5pt}
\caption{Combined fit result for disconnected contribution $G^s_M(Q^2=0.0515\, \text{GeV}^2)$ with $m_\pi = 207 \,\text{MeV}$. The bands show fits to the 3pt/2pt ratios. The current insertion time $t_1$ is shifted by half the sink-source separation for clarity.}
\la{fig:Ratio}
\end{figure}
The unprecedented precision we obtain in statistics is partly due to the fact that we calculate the low-mode contribution to the loop exactly without any stochastic noise. We find that about $15\%-25\%$ of the signal is saturated by the low modes while determining the $s$-quark matrix elements in this calculation. 
 
Next, we explore the $Q^2$ dependence of $G^s_M(Q^2)$ to obtain the strange magnetic moment at $Q^2=0$. We compare both the dipole form~\cite{Hand:1963zz} and the model independent $z$-expansion fit~\cite{Hill, Epstein} given by
\bea \la{zexp}
G^{s,z-exp}_{M}(Q^2)\! =\! \sum^{k_\text{max}}_{k=0}\! a_k z^k \!&,&\!z\!=\!\frac{\sqrt{t_{\text{cut}}+Q^2}-\sqrt{t_{\text{cut}}}}{\sqrt{t_{\text{cut}}+Q^2}+\sqrt{t_{\text{cut}}}}.
\eea 
We set $t_{\text{cut}}= (2m_K)^2$. We keep the first three coefficients multiplying $z^k$ in the $z$-expansion formula and perform fits versus $Q^2$. We calculate the jackknife ensemble average $a_{2,avg}$ of the coefficient $a_2$ and then perform another fit by setting $a_2$ centered at $a_{2,\text{avg}}$ with a prior width equal to $2\times |a_{2, \text{avg}}\vert$. We find the effect of setting this prior is almost insignificant for the 24I and 32I ensemble data, especially at heavier quark masses. However, the prior stabilizes the extrapolation of $G^s_M(Q^2)$ for pion masses around the physical point for the 48I ensemble. Since the $z$-expansion method guarantees that $a_k$ coefficients are bounded in size and that higher order $a_k$'s are suppressed by powers of $z^k$, we carefully check the effect of the $a_3$ coefficient in our fit formula and estimate this effect to calculate the systematic uncertainties in the $z-\text{expansion}$ fit. We present the extrapolation of $G^s_M(0)$ using both the dipole and $z$-expansion methods in Fig.~\ref{fig:Qdep} with the smallest lattice spacing $\mathnormal{a} = 0.0828(3)\, \text{fm}$ used in our simulation and lattice data at the unitary point for the 32I ensemble with a pion mass $m_\pi = 330\, \text{MeV}$. The present calculation does not provide any conclusive evidence of any statistically significant difference between these two methods, as seen in the figure. However, because of model independence and goodness of the fit, we use $z-\text{expansion}$ fit results in the rest of our calculation. 

\begin{figure}[htbp]
\begin{center}
\includegraphics[height=4.0cm,width=7.0cm]{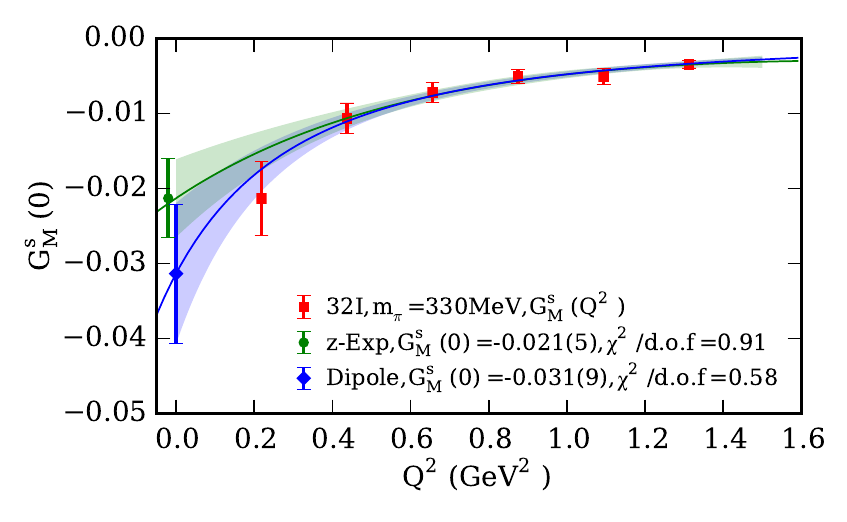}
\end{center}
\setlength\abovecaptionskip{-9pt}
\setlength\belowcaptionskip{-9pt}
\caption{Comparison between the classical dipole form and the model-independent $z$-expansion fit to study the $Q^2$ dependence of $G^s_M$ and extract $G^s_M(0)$. The $G^s_M(Q^2)$ data points correspond to the 32I ensemble with quark masses corresponding to $m_\pi = 330\, \text{MeV}$.}
\la{fig:Qdep}
\end{figure}

From the $z$-expansion extrapolations, we obtain $24$ different estimates of $G^s_M(0)$ from four different lattice ensembles with varying quark masses. As the nucleon 2pt correlation function depends on the valence quark masses and the strange quark matrix elements depend on $m_{\text{loop}}$, we use a chiral extrapolation linear in $m_\pi$ and $m_{\text{loop}}=m_K$~\cite{Musolf, Hemmert1, Hemmert2, Chen}. To account for the partial quenching effect with the valence-sea pion mass ($m_{\pi, vs}$), and the $\mathcal{O}(a^2)$ correction and volume dependence~\cite{Beane}, the global fit formula we use for the extrapolation of $G^s_M(0)$ to the physical point is
\bea \la{gsmfit}
&&G^s_M(0; m_{\pi}, m_{\pi,vs}, m_K, a , L) = A_0 + A_1 m_\pi + A_2 m_K  \nn \\
&&+ \!A_3m^2_{\pi,vs} \!+ A_4 a^2\! +A_5 m_{\pi} \bigg(1-\frac{2}{m_{\pi} L}\bigg)\!e^{-m_{\pi} L },
\eea

\begin{figure}
\begin{center}
\includegraphics[height=4.5cm,width=7.5cm]{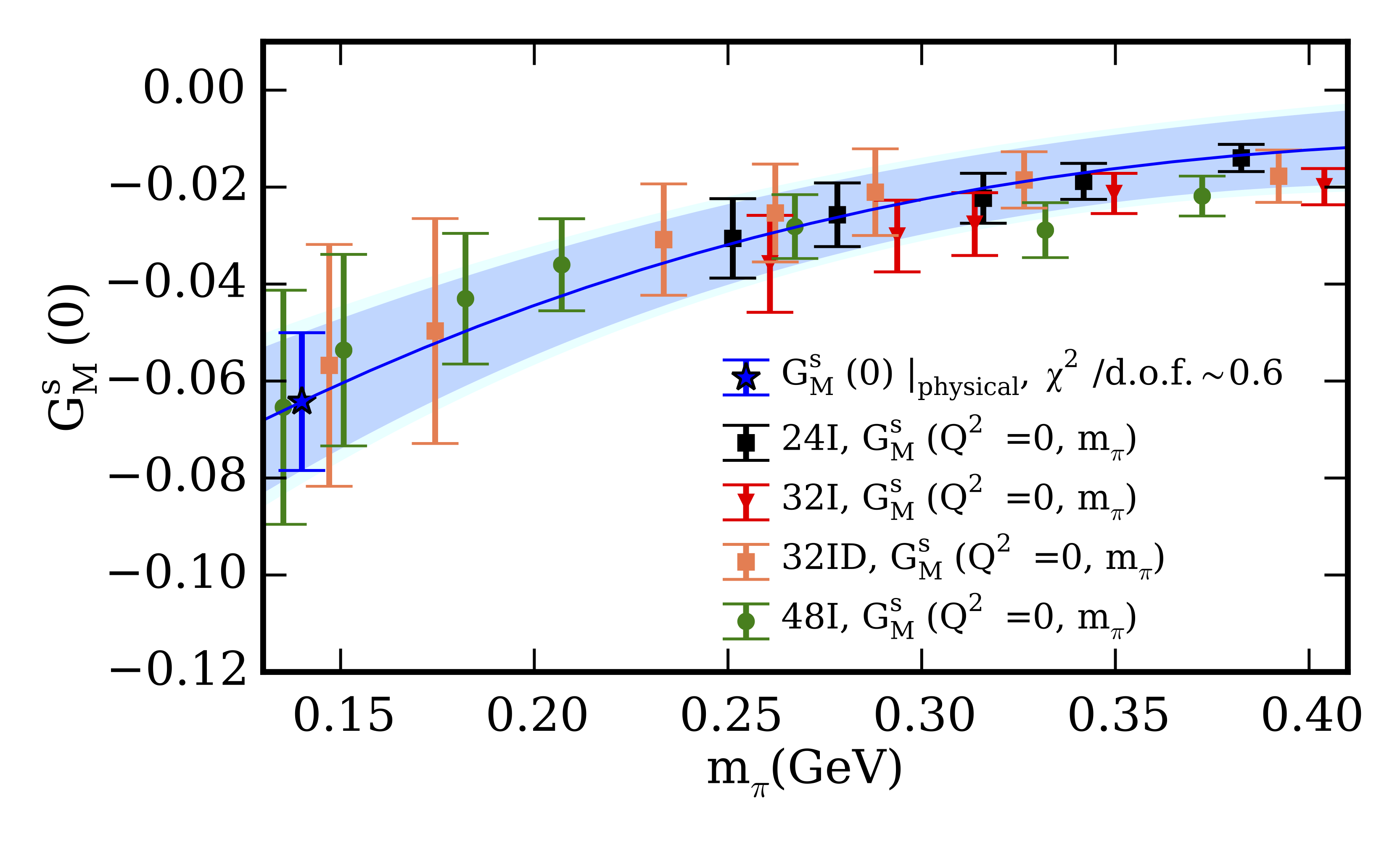}
\end{center}
\setlength\abovecaptionskip{-9pt}
\setlength\belowcaptionskip{-7pt}
\caption{Strange magnetic moment at 24 quark masses on 24I, 32I, 48I, and 32ID ensembles as a function of the pion mass. The curved blue line in the figure shows the behavior in the infinite volume and continuum limit. The cyan band shows the combined statistical and systematic uncertainties added in quadrature.}
\la{fig:continuum}
\end{figure}
where $m_{\pi}\,(m_K)$ is the valence pion (kaon) mass and $m_{\pi, vs}$ is
the partially quenched pion mass {\mbox{$m_{\pi, vs}^2 = 1/2(m_{\pi}^2 + m_{\pi, ss}^2)$}} with $m_{\pi, ss}$ the pion mass corresponding to the sea quark mass. $A_4$ includes the 
mixed action parameter $\Delta_\text{mix}$~\cite{Lujan:2012wg}. The extrapolation of the strange magnetic moment is shown in Fig.~\ref{fig:continuum} and at the physical point in the limit $a\to 0$ and $L\to\infty$ we obtain
\bea
G^s_M(0)\vert_{\text{physical}} = - 0.064(14)(04)(06)(06) \,\mu_N. 
\eea 
Here, the uncertainties in the parentheses are from the statistics, interpolation to the physical $s$-quark mass~\cite{Yibo2}, introducing $a_3$ coefficients in the $z$-expansion fit, and the global fit formula for the continuum extrapolation of $G^s_M(0)$, respectively. To calculate the uncertainty associated with the global fit formula, we consider the higher order volume correction terms $(m^{3/2}_\pi / \sqrt{L})e^{-m_{\pi}L}$~\cite{Beane}, $m_N m_K$~\cite{Hemmert2}, $\log{m_{\pi}^2}$, and $m_{\pi,vs}$. We obtain the fit coefficients: $A_1=0.61(16)$, $A_2=-2.26(49)$, $A_3=0.31(12)$, $A_4=0.015(16)$, and $A_5=-4.0(2.4)$ with the sign of $A_5$ consistent with that in Ref.~\cite{Beane}. We note that the $\mathcal{O}(a^2)$ effect is small, whereas the partial quenching effect and the volume correction along with the quark mass dependence play roles in our global fit. While $G^s_M(0)$ values for different ensembles are consistent within uncertainty near $m_\pi=250\,\text{MeV}$, from the fit coefficients it can be seen that, near $m_\pi=400\,\text{MeV}$, $G^s_M(0)$ calculated from the 48I ensemble is more negative due to the partial quenching effect.

 For a given valence quark mass we fit $G^s_E(Q^2)$ using the $z-$expansion method described above and calculate the charge radius from the fitted slope of the data using the definition $\langle r_s^2\rangle_E \equiv -6 \frac{dG^s_E}{dQ^2}\vert_{Q^2=0}$. The net strangeness in the nucleon is zero, and thus $G^s_E (0) = 0$, which we confirm in our simulation. Chiral extrapolation to the $\langle r_s^2\rangle_E$ data is obtained from Ref.~\cite{Hemmert2}. Because the method of finite volume correction of nucleon charge radius is less clear and hard to obtain~\cite{Hall, Tiburzi}, we employ an empirical formula for the volume correction to describe our lattice data. The empirical fit formula we use to obtain $\langle r_s^2\rangle_E$ at the physical point is
\bea \la{rse}
&&\langle r_s^2\rangle_E  (m_{\pi}, m_{\pi,vs}, m_K, a, L)= A_0 \!+\! A_1\log{(m_K)} \! \nn \\
&&+\! A_2 m^2_{\pi}+ A_3m^2_{\pi,vs} + A_4 a^2 + A_5 \sqrt{L} e^{-m_{\pi} L}.
\eea
We find that the volume correction term similar to the pion charge radius term derived in Ref.~\cite{Tiburzi} describes our lattice data well. From the fitted values of the coefficients in Eq.~(\ref{rse}), namely, $A_1=0.03(2)$, $A_2=-0.04(8)$, $A_3=0.03(2)$, $A_4=-0.0004(27)$, and $A_5=0.001(7)$, it is seen that among different contributions the quark mass dependence and partial quenching effect are more important in determining $\langle r_s^2\rangle_E$ from our lattice data. We also consider $e^{-m_\pi L}$, $m_K$ instead of $\log{m_K}$, $1/m_N^2$~\cite{Hemmert2}, $m_{\pi,vs}$ and calculate a systematic error derived from different terms in the global fit formula. We present the value of $\langle r_s^2\rangle_E$ at the physical point in Fig.~\ref{fig:Radius} which gives
\bea \la{radius}
\langle r_s^2\rangle_E \vert_{\text{physical}} = -0.0043 (16)(02)(08)(07)\, \text{fm}^2.
\eea
\begin{figure}
\begin{center}
\includegraphics[height=4.5cm,width=7.5cm]{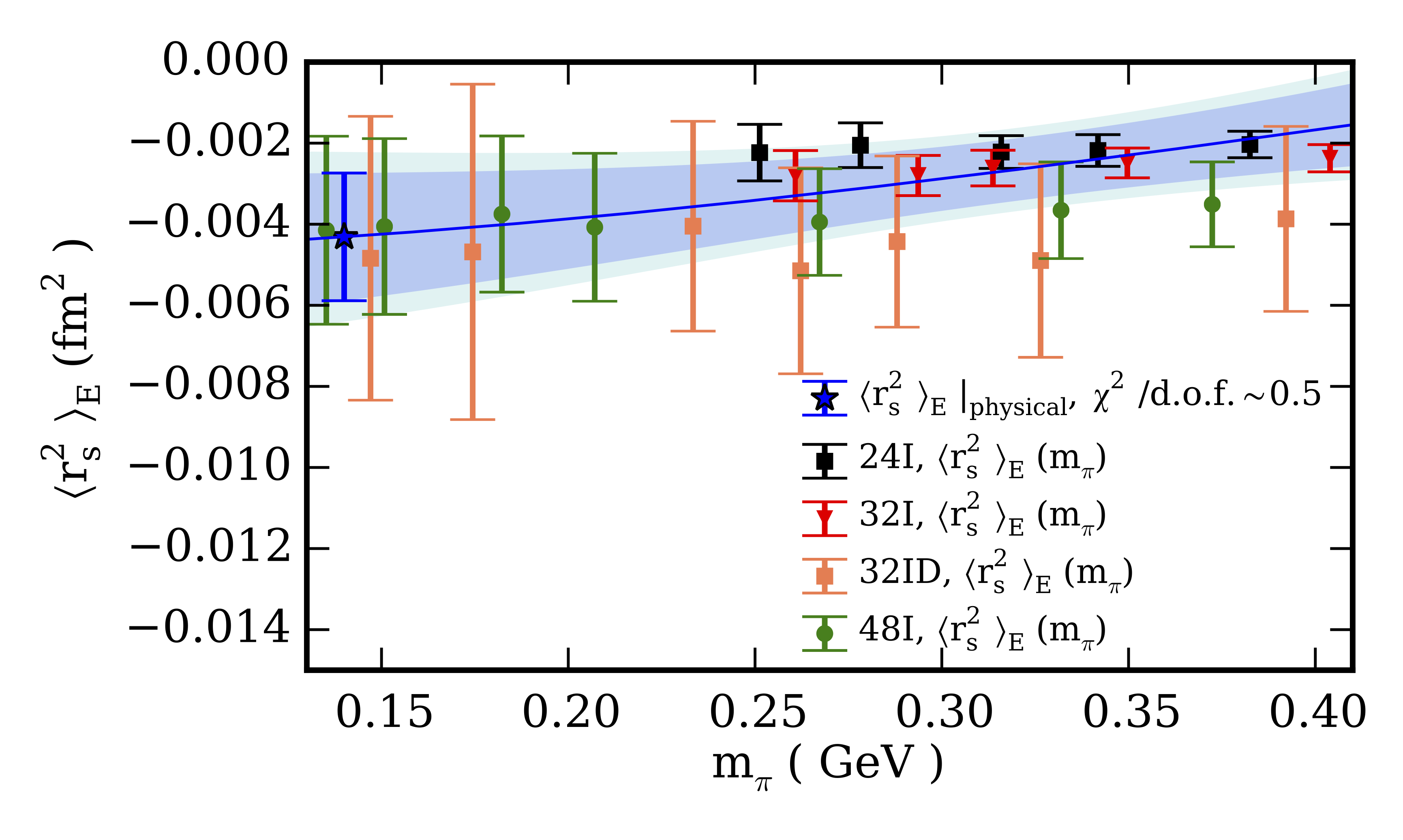}
\end{center}
\setlength\abovecaptionskip{-9pt}
\setlength\belowcaptionskip{-7pt}
\caption{Strange charge radius at 24 quark masses on 24I, 32I, 48I, and 32ID ensembles as a function of the pion mass. The curved blue line in the figure shows the behavior in the infinite volume and continuum limit. The cyan band shows the combined statistical and systematic uncertainties added in quadrature.}
\la{fig:Radius}
\end{figure}
The uncertainties in the second and third parentheses of Eq.~(\ref{radius}) are obtained using similar methods described in the case of $G^s_M(0)$. The lowest $Q^2$ values for 48I and 32ID ensembles are 0.051 and 0.073 $\text{GeV}^2$ respectively, which are almost $3-4$ times smaller than the lowest $Q^2=0.22\,\text{GeV}^2$ of the 24I and 32I ensemble. As extracting the charge radius from the FF data can be sensitive to the lowest available $Q^2$, this can affect our determination of $\langle r_s^2\rangle_E $. A $20\%$ uncertainty in introducing the $a_3$ term in the $z-$expansion has been included as a systematic in the final result of $\langle r_s^2\rangle_E $.

Finally, we present Fig.~\ref{fig:comparison} to compare our result of $G^s_M(0)$ and $G^s_M(Q^2 = 0.1 \,\text{GeV}^2) = -0.037(10)(05)$ with some other measurements of $G^s_M(0)$ and global analyses of $G^s_M$ at \mbox{$Q^2=0.1 \,\text{GeV}^2$}. We strongly believe that controlling excited-state contamination, performing the simulation near the physical pion mass, and considering the finite size effect altogether play an important role in determining the strange magnetic moment as observed in our lattice simulation.
\begin{figure}
\begin{center}
\includegraphics[height=6cm,width=8.0cm]{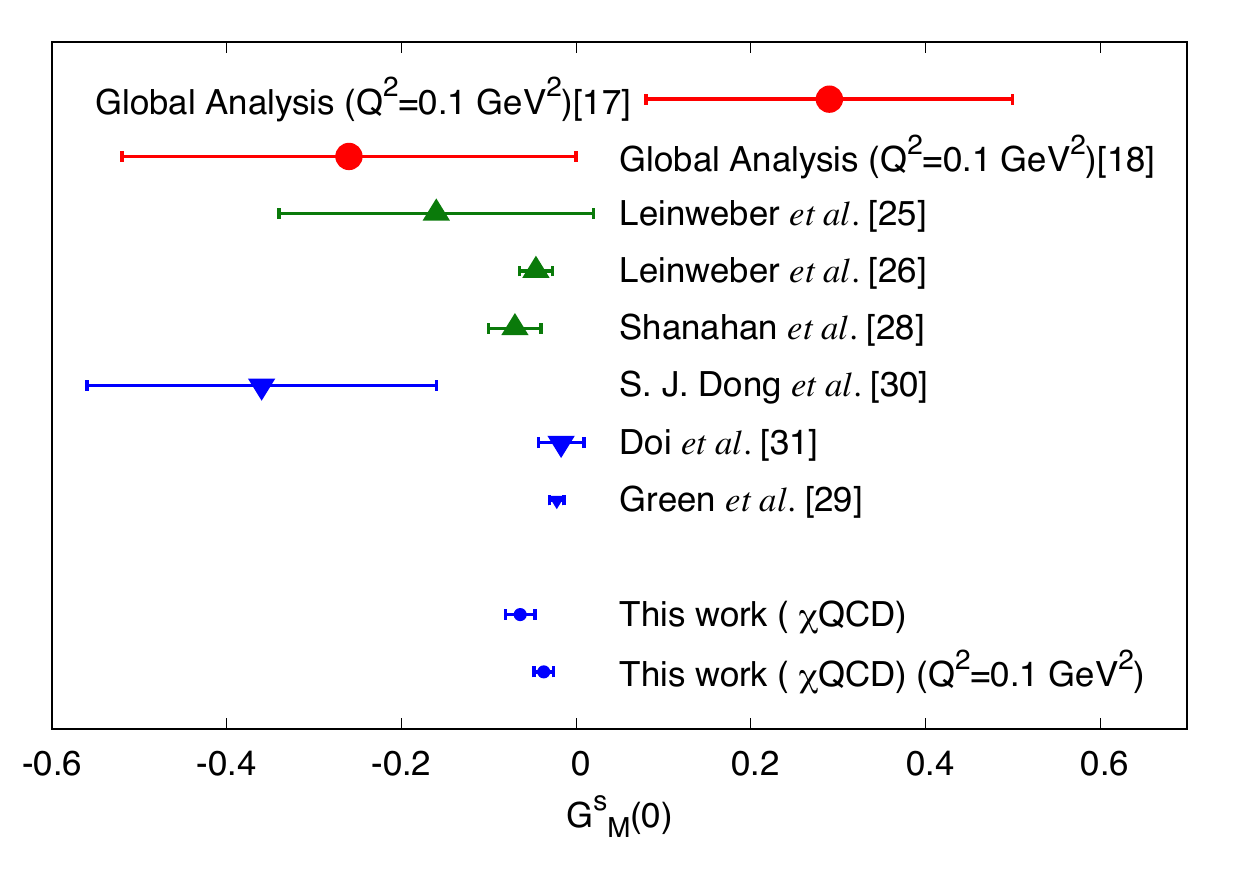}
\end{center}
\setlength\abovecaptionskip{-9pt}
\setlength\belowcaptionskip{-7pt}
\caption{Comparison of some of the many determinations of the strange magnetic moment. Results in {\emr red} are from the global analysis of world data, results in {\color{green!90!blue}green} are from indirect calculations, and results in {\emb blue} are from lattice QCD calculations. }
\la{fig:comparison}
\end{figure}

In conclusion, we have performed a robust first-principles lattice QCD calculation using four different $2+1$ flavor dynamical fermion lattice ensembles including, for the first time, the physical pion mass to explore the quark mass dependence and with finite lattice spacing and volume corrections to determine the strange quark matrix elements in the vector channel. We have performed a two-state fit where we combined both the ratio method and the summed-ratio method to control excited-state contamination. The statistical error is greatly reduced by improving the nucleon propagator with LMS and quark loop with LMA. To explore the strange vector form factors at different momentum transfers, we implemented model-independent $z-$expansion fits. Given our precise lattice prediction for the strange quark magnetic moment of $G^s_M(0) = - 0.064(17) \mu_N$ and strange charge radius $\langle r_s^2\rangle_E = -0.0043(21) \, \text{fm}^2$ at the physical point with systematic errors included, we anticipate these results to be verified by experiments in the future and, together with experimental inputs, to lead to a more precise determination of various weak form factors. 

%%%%%%%%%%%%%%%%%%%%%%%%%%%%%%%%%%%%%%%%%%%
%%%%%%%%%%%%%%%%%%%%%%%%%%%%%%%%%%%%%%%%%%%%
%%%%%%%%%%%%%%%%%%%%%%%%%%%%%%%%%%%%%%%%%%%
%%%%%%%%%%%%%%%%%%%%%%%%%%%%%%%%%%%%%%%%%%%

\begin{acknowledgments}
  \textit{Acknowledgments:} We thank the RBC and UKQCD Collaborations for providing their DWF gauge configurations. This work is supported in part by the U.S. DOE Grant No. DE-SC0013065. This research used resources of the Oak Ridge Leadership Computing Facility at the Oak Ridge National Laboratory, which is supported by the Office of Science of the U.S. Department of Energy under Contract No. DE-AC05-00OR22725. A.A. is supported by the NSF CAREER Grant No. PHY-1151648 and in part by the DOE Grant No. DE-FG02-95ER-40907.
\end{acknowledgments}

\providecommand{\href}[2]{#2}
\begingroup\raggedright

\endgroup

\end{document}